\documentclass[lettersize,journal]{IEEEtran}

\usepackage{amsmath,amsfonts}
\usepackage{algorithmic}
\usepackage{algorithm}
\usepackage{array}
\usepackage[caption=false,font=normalsize,labelfont=scriptsize,textfont=scriptsize]{subfig}  
\usepackage{textcomp}
\usepackage{stfloats} 
\usepackage{url}
\usepackage{verbatim}
\usepackage{graphicx}
\usepackage{enumitem}
\usepackage{booktabs} 
\usepackage{bm}
\usepackage{bbding}
\usepackage{mathrsfs} 
\usepackage{amsthm} %
\usepackage[colorlinks=true,linkcolor=black,citecolor=black,urlcolor=blue,]{hyperref}
\usepackage{amssymb}
\usepackage{cite}

\begin{document}
\title{Environment-Aware Codebook for Reconfigurable Intelligent Surface-Aided MISO Communications}
\author{Xing Jia,  Jiancheng An,~\IEEEmembership{Member,~IEEE},  Hao Liu, Hongshu Liao, Lu Gan,  and Chau Yuen,~\IEEEmembership{Fellow,~IEEE}
\thanks{This work was partially supported by Sichuan Science and Technology Program under Grant 2023YFSY0008 and 2023YFG0291. This research is supported by the Ministry of Education, Singapore, under its MOE Tier 2 (Award number MOE-T2EP50220-0019). Any opinions, findings and conclusions or recommendations expressed in this material are those of the author(s) and do not reflect the views of the Ministry of Education, Singapore.}
\thanks{Xing Jia, Hao Liu, Hongshu Liao and Lu Gan are with the School of Information and Communication Engineering, University of Electronic Science and Technology of China (UESTC), Chengdu 611731, China, and also with the Yibin Institute of UESTC, Yibin 644000, China (e-mail: xingjia1999@163.com; liu.hao@std.uestc.edu.cn; hsliao@uestc.edu.cn; ganlu@uestc.edu.cn). Jiancheng An is with the Engineering Product Development (EPD) Pillar, Singapore University of Technology and Design (SUTD), Singapore 487372 (e-mail: jiancheng\_an@163.com). C. Yuen is with the School of Electrical and Electronics Engineering, Nanyang Technological University (NTU), Singapore 639798 (e-mail:
chau.yuen@ntu.edu.sg).}
}
\maketitle

\begin{abstract}
Reconfigurable intelligent surface (RIS) is a revolutionary technology that can customize the wireless channel and improve the energy efficiency of next-generation cellular networks. This letter proposes an environment-aware codebook design by employing the statistical channel state information (CSI) for RIS-assisted multiple-input single-output (MISO) systems. Specifically, first of all, we generate multiple virtual channels offline by utilizing the location information and design an environment-aware reflection coefficient codebook. Thus, we only need to estimate the composite channel and optimize the active transmit beamforming for each reflection coefficient in the pre-designed codebook, while simplifying the reflection optimization substantially. Moreover, we analyze the theoretical performance of the proposed scheme. Finally, numerical results verify the performance benefits of the proposed scheme over the cascaded channel estimation and passive beamforming as well as the existing codebook scheme in the face of channel estimation errors, albeit its significantly reduced overhead and complexity. 
\end{abstract}

\begin{IEEEkeywords}
Reconfigurable intelligent surface (RIS), codebook design, channel estimation, reflection optimization.
\end{IEEEkeywords}

\section{Introduction}
\IEEEPARstart {R}{ecently}, reconfigurable intelligent surface (RIS) has gained widespread attention for improving system energy efficiency in a low-cost manner \cite{ref1_AO,an_equi}. Generally,  RIS is a programmable metasurface composed of numerous passive elements, and each element can independently adjust the phase/amplitude of incident signals \cite{ref2_C1}. Furthermore, RIS can operate in a full-duplex mode while circumventing the common self-interference issue and can be directly integrated with the existing wireless network to perform its function \cite{xr3}. Thanks to the aforementioned benefits, RIS is envisioned as a critical technology enabling future wireless networks \cite{xr3}.

In order to unlock the full potential of RIS-assisted communication systems, accurate channel state information (CSI) is generally demanded, which poses a big challenge due to the large number of passive elements on RIS. To address this issue, an ON/OFF scheme was proposed in \cite{ref2_C1} to estimate the cascaded base station (BS)-RIS-user channels. {Moreover, a discrete Fourier transform (DFT)-based phase shifting configuration was adopted in \cite{Gua_HY} to perform uplink cascaded channel estimation.  Furthermore, the authors of \cite{zhangSB,xc3} proposed deep learning-based channel estimation methods to reduce the training overhead by exploiting the sparsity of the beamspace.}

{In addition, joint active beamforming and reflection coefficient optimization is generally non-trivial. Recently, the authors of \cite{ref1_AO} proposed an alternate optimization (AO) algorithm to design the precoding and reflection coefficient (RC) matrices for minimizing the transmit power at the BS.} Furthermore, a block coordinate descent (BCD) algorithm was applied in \cite{ref8_R3} to implement RC optimization for maximizing the average sum-rate. The authors of \cite{STA_CSI} designed the RC  based on the statistical CSI for maximizing spectral efficiency. Moreover, for minimizing the transmit power, the authors of  \cite{xr1,xr3} proposed a gradient descent algorithm to optimize the RC.

{Nevertheless, the existing joint optimization schemes generally require high complexity and excessive  overhead to probe the reflected channels. Recently, a novel codebook scheme that obtains a suboptimal solution in a low complexity was investigated in \cite{ref14_WCL,an_oujld, an_equi,Gua_HY}. Specifically, the authors of \cite{ref14_WCL} investigated a random codebook scheme, while  \cite{an_equi} and \cite{an_oujld} proposed a sum Euclidean distance maximizing codebook by considering the discrete and continuous phase shifts, respectively.} In addition, the limited feedback bit allocation scheme based on channel structure for adapting to diverse environments was investigated in \cite{ref10_M2}. Moreover, an adaptive codebook design relying on a deep neural network was proposed by considering the limited feedback link \cite{kim}.

Note that the existing  schemes generally designed the codebook without considering the wireless channel  \cite{an_oujld,an_equi}. We propose a novel environment-aware codebook to address this issue by exploiting the statistical CSI. Considering the low communication rate of the feedback link, the RIS configuration is completed only by feeding back the optimal index. On one hand, the proposed scheme can strike a beneficial trade-off between training overhead and system performance. On the other hand, the designed codebook based on  statistical CSI attains performance improvements compared to the existing codebook schemes. In addition, we analyze the theoretical performance and  complexity of the proposed scheme. Finally, simulation results verify the performance benefits of the proposed scheme and its robustness against channel estimation errors.

\section{System and Channel Models}\label{SystemModel}
\subsection{System Model}
\begin{figure}[!t]
\centering
\includegraphics[width=2.2in]{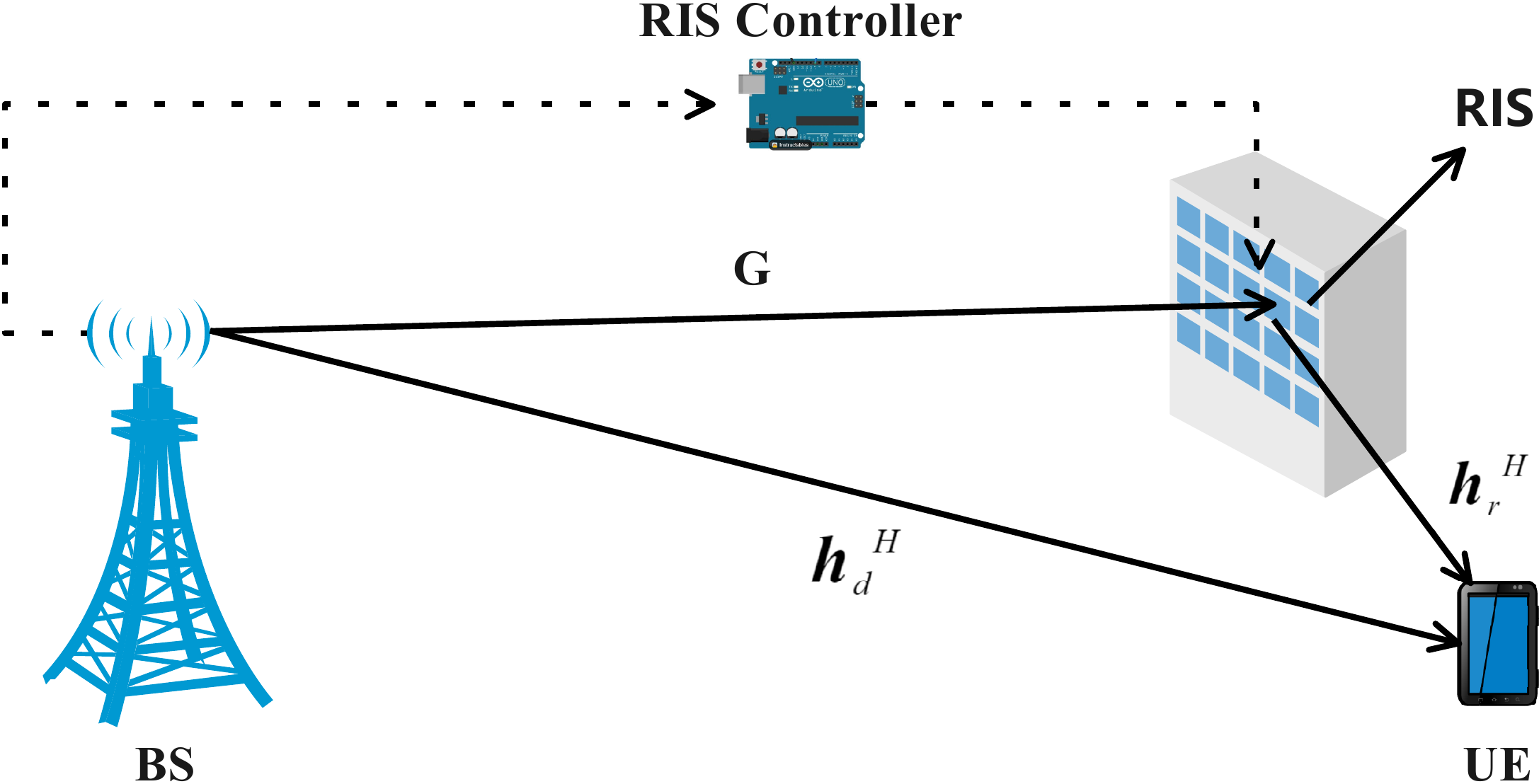}
\caption{An RIS-assisted downlink MISO system.}
\vspace{-0.6cm} 
\label{fig_1}
\end{figure}

As shown in Fig. \ref{fig_1}, we consider an RIS-assisted multiple-input single-output (MISO) communication system in a single cell. The RIS having $N$ reflecting elements assists in the downlink communication from a BS with $M$ antennas to a single-antenna user equipment (UE). Let $\mathcal{N}=\{1,2,...,N \}$ and  $\mathcal{M}=\{1,2,...,M \}$, respectively. Moreover, all RIS elements are connected to a smart RIS controller, which is capable of independently adjusting the phase of the incident signals. In addition, considering the practical hardware implementation, we assume the discrete phase shift of uniformly quantifying $[0,2\pi)$ imposed by each RIS  element. Let $b$ denote the number of quantization bits associated with each element. Thus, we have RC set $\mathscr{E} = \{ {e^{j0}},{e^{j\Delta \theta }},...,{e^{j({2^b} - 1)\Delta \theta }}\}$, where $\Delta \theta = 2\pi /{2^b}$ denotes the quantified interval. Let $ \boldsymbol{\phi}= [{\varphi _1}, {\varphi _2},...,{\varphi _N}]^T \in {\mathbb{C}^{N \times 1}}$ denote the RC vector, where $\varphi _n \in \mathscr{E}$, ${n}\in\mathcal{N}$ denotes the RC of the $n$-th RIS element. 

Let $\mathbf{\Theta} = \text{diag}(\boldsymbol{\phi})\in {\mathbb{C}^{N \times N}}$ denote the reflection matrix of the RIS. This letter assumes the narrow-band flat fading channel and the time-division duplex (TDD) mode. Let $\mathbf{G} \in {\mathbb{C}^{N \times M}}$, ${\mathbf{h}^H_\text{r}} \in {\mathbb{C}^{1 \times N}}$, and ${\mathbf{h}^H_\text{d}} \in {\mathbb{C}^{1 \times M}}$ denote the complex baseband channel  from the BS to the RIS, from the RIS to the UE, and from the BS to the UE, respectively. As a result, the downlink channel $\mathbf{h}^H \in {\mathbb{C}^{1 \times M}}$ can be denoted as $\mathbf{h}^H={ {\mathbf{h}^{H}_\text{d}} + {\mathbf{h}^H_\text{r}}{\mathbf{\Theta}}{\mathbf{G}}} = \mathbf{h}^{H}_\text{d} +  \boldsymbol{\phi}^T \mathbf{D}$, where $\mathbf{D}=\text{diag}(\mathbf{h}^H_\text{r})\mathbf{G} \in \mathbb{C}^{N \times M}$ denotes the cascaded BS-RIS-UE channel.

First, we consider the uplink channel estimation process. Let ${x}$ denote the normalized pilot signal, satisfying $\mathbb{E} \left\{{{{\left| {{x}} \right|}^2}} \right\} = 1$, and ${p_\text{u}}$ denotes the average power of pilot signal. {Assume the system having $T$ training time slots (TSs)and define $\mathcal{T} = \{1,2,...,T\}$. Thus, the BS received signal at $t$-th TS  ${\mathbf{y}_t} \in {\mathbb{C}^{M \times 1}}$ is written as
\begin{equation}
\label{testHH}
{{\mathbf{y}_t} =\mathbf{h}_{t}\sqrt {{p_\text{u}}} { {x} + {\mathbf{n}_{\text{u},t}}}= \left({\mathbf{h}_\text{d}} + {\mathbf{D}^H}\boldsymbol{\phi}_t^*\right )}\sqrt {{p_\text{u}}} { {x} + {\mathbf{n}_{\text{u},t}}}, 
\end{equation}where ${\forall}{t}\in\mathcal{T}$, ${\mathbf{n}_{\text{u},t}} \in {\mathbb{C}^{M \times 1}}$ and ${\phi}_{t}$ denote the additive white Gaussian noise (AWGN) at the BS with the average noise power of ${\sigma^2 _\text{u}}$, satisfying ${\mathbf{n}_\text{u}} \sim\mathcal{CN}(\mathbf{0},{\sigma^2_\text{u}}{\mathbf{I}_M})$ and the RIS RC vector at the $t$-th TS.

Next, we consider the downlink communications. The total power at the BS is denoted by ${p_\text{d}}$, and ${z}$ represents the  signal transmitted from the BS. Thus, the signal received at the UE is denoted as 
\begin{equation}
\label{2}
{r}_{t} =\mathbf{h}_{t}^H \sqrt {{p_\text{d}}} {\mathbf{w}{z} + {{n}_{\text{d},t}}}= \left(\mathbf{h}^{H}_\text{d} +  \boldsymbol{\phi}_{t}^T \mathbf{D}\right) \sqrt {{p_\text{d}}} {\mathbf{w}{z} + {{n}_{\text{d},t}}},
\end{equation}where $t \in [T+1,L]$, $L$ denote coherent time slot, $\mathbf{w} \in {\mathbb{C}^{M \times 1}}$ denotes the transmit beamforming vector, satisfying ${\mathbf{\left \|w\right \|}^2} = 1$, and ${{n}_{\text{d},t}} \sim \mathcal{CN}(0,{\sigma^2 _\text{d}})$ denotes the AWGN at the UE with the average noise power of ${\sigma^2 _\text{d}}$ at the $t$-th TS.}

\subsection{Channel Model}
Before going on further, we elaborate on the Rician channel model adopted in this letter. Specifically,  the direct channel ${\mathbf{h}_\text{d}}$ is generated by
\begin{equation}
\label{generate_ch}
{\mathbf{h}_\text{d}} = \sqrt {{\beta _\text{d}}/\left({K_\text{d}} + 1\right)} \left(\sqrt {{K_\text{d}}} {\mathbf{h}^\text{LoS}_\text{d}} + {\mathbf{h}^\text{NLoS}_\text{d}}\right),
\end{equation}where ${\mathbf{h}^\text{LoS}_\text{d}}$ and ${\mathbf{h}^\text{NLoS}_\text{d}}$ represent the line-of-sight (LoS) and non-LoS (NLoS) components of ${\mathbf{h}_\text{d}}$, respectively. {The NLoS component of ${\mathbf{h}_\text{d}}$ is modeled by complex Gaussian distribution, i.e., ${\mathbf{h}^\text{NLoS}_\text{d}}\sim \mathcal{CN}(\mathbf{0},{\mathbf{I}_M})$; ${\beta _\text{d}}$ and ${K_\text{d}}$  denote the  path loss and  Rician factor of the direct channel ${\mathbf{h}_\text{d}}$, respectively. Similarly, $\mathbf{G}$ and ${\mathbf{h}_\text{r}}$ are generated by  a similar way as (\ref{generate_ch}).} Furthermore, we consider  a uniform linear array (ULA) of $M$ antennas at the BS. Let $\mathbf{a}_{\text{BS}} \in {\mathbb{C}^{M \times 1}}$ denote the steering vector of the BS, and the $m$-th entry of the $\mathbf{a}_{\text{BS}}$ is denoted as $ {{{e}}^{j2\pi (m - 1){d_\text{BS}}\sin {\phi }/\lambda }},\ {\forall} {m}\in\mathcal{M}$, where ${d_\text{BS}}$ denotes the antenna spacing of the BS, $\lambda $ denotes the wavelength of the signal, and $\phi \in [-\pi/2,\pi/2)$ denotes the angle of departure (AoD) or the angle of arrival (AoA). 

Furthermore, we model the RIS as a uniform planar array (UPA). Let  ${{\mathbf{a}}_\text{R}}(\alpha ,\gamma ) \in {\mathbb{C}^{N \times 1}}$ denote the steering vector of the RIS. Specifically, the $n$-th entry of  $\mathbf{a}_{\text{R}}$ is denoted as \\$ {e^{j2\pi {d_\text{R}}\sin {\gamma}[\lfloor {  \frac{n-1}{N_\text{x}} } \rfloor \sin{\alpha} + ( ( n-1) - {\lfloor {  \frac{n-1}{N_\text{x}} } \rfloor} {N_\text{x}})\cos {\alpha }]/\lambda }},\ {\forall} {n}\in\mathcal{N}$, where $d_\text{R}$ denotes the element spacing of the RIS, and ${N_\text{x}}$ denotes the number of reflecting elements deployed at each row \cite{an_oujld}. Moreover, $\alpha \in [0,\pi)$ and $\gamma \in [ - \pi/2,\pi /2)$ denote the azimuth and elevation AoA/AoD, respectively.  As a result, the LoS components of the ${\mathbf{G}}$, ${\mathbf{h}_\text{r}}$, and ${\mathbf{h}_\text{d}}$ can be denoted as ${{\mathbf{a}}_\text{BS}}{(\phi _\text{d}^\text{A})}$, $ { {\mathbf{a}}_\text{R}}(\alpha_\text{g}^\text{A},\gamma_\text{g}^\text{A}){{\mathbf{a}}_\text{BS}}{(\phi_\text{g}^\text{D})^H}$, and ${{\mathbf{a}}_\text{R}}{(\alpha _{\text{r}}^\text{A},\gamma _{\text{r}}^\text{A})}$, respectively, where $\phi _\text{d}^\text{A}$ and $\phi _\text{g}^\text{D}$ denote the AoA from the UE to the BS and the AoD from the BS to the RIS, respectively; $\alpha_\text{g}^\text{A}$ and $\gamma_\text{g}^\text{A}$ denote the azimuth and elevation AoA from the BS to the RIS, respectively; $\alpha_{\text{r}}^\text{A}$  and $\gamma_{\text{r}}^\text{A}$ denote the azimuth and elevation AoA from the UE to the RIS, respectively. 
\section{The Proposed Codebook-Based Design} \label{proposed}
{In this section, we first present the protocol of RIS-assisted system based on codebooks in Section \ref{codebook}, then we elaborate on the steps of our environment-aware codebook design in Section \ref{mycodebook}. Finally, we analyze the advantages of the proposed scheme in Section \ref{Advantages}.}
\subsection{The Proposed Protocol} \label{codebook}
\subsubsection{Environment-Aware Codebook Generation} The codebook scheme could significantly reduce the implementation complexity and backhaul consumption \cite{an_oujld}. In this letter, we first design an environment-aware codebook consisting of a set of RCs according to the statistical CSI, based on which we pursue a suboptimal solution. Thus the BS only needs to feedback the optimal index to the RIS through the control link. The detailed design of the environment-aware codebook will be elaborated in Section \ref{mycodebook}.

\subsubsection{Composite Channel Estimation} {In sharp contrast to the existing channel estimation schemes that estimate the direct and cascaded channels separately, we directly estimate the superimposed end-to-end channel spanning from the UE to the BS for each given RIS RC in the pre-designed codebook, which requires $T$ training slots. Specifically, the composite channel is denoted as ${{\mathbf{h}_{ t}} = {\mathbf{h}_\text{d}} + {\mathbf{D}^H}{\boldsymbol{\phi}_t ^ *}}$ at the $t$-th training TS. By applying the least square (LS) \cite{an_equi} and minimum mean square error (MMSE) \cite{Gua_HY} technique, we could estimate the composite channel as
\begin{align} 
{\hat{\mathbf{h}}_{t,\text{LS}}} &= {x^*}{\mathbf{y}_{t}}/\sqrt {{p_\text{u}}},  \quad {\forall}{t}\in\mathcal{T},\label{xxx}\\
{{\hat{\mathbf{h}}_{t,\text{MMSE}}}}&= \mathbf{R}_{h} \left(\mathbf{R}_{h}^H + {\sigma^2_\text{u}}{\mathbf{I}_M}/{p_\text{u}}\right)^{-1}{\hat{\mathbf{h}}_{t,\text{LS}}},  \quad {\forall}{t}\in\mathcal{T},\label{xxx1}
\end{align}where $\mathbf{R}_{h}=\mathbb{E}\left\{\mathbf{h}\mathbf{h}^H\right\}$ denotes  channel correlation matrix of $\mathbf{h}$. Note that the proposed composite channel estimation scheme has significant advantages over the scheme of \cite{ref2_C1} in terms of implementation complexity.}

\subsubsection{Downlink Transmit Beamforming} {According to the channel estimation in (\ref{xxx}), (\ref{xxx1}) and by applying the channel reciprocity in TDD mode \cite{tangWKl}, at the $t$-th training TS, ${t}\in\mathcal{T}$. The active transmit beamforming vector ${\hat{\mathbf{w}}_t}$ can be obtained by applying the maximum ratio transmission criterion  \cite{ref14_WCL} as $\mathop {{\hat{\mathbf{w}}_t}}\limits  = \mathop {{\hat{\mathbf{h}}_{t}}}\limits / || {\mathop {{\hat{\mathbf{h}}_{t}}}\limits } || = {x^*}{\mathbf{y}_{t}}/ || {\mathop {{{\mathbf{y}}_{t}}}\limits } ||$.} Therefore, at the $t$-th training TS, the achievable downlink rate can be expressed as
\begin{equation}
\label{5}
{R_t} = {\log _2}\left(1 + \frac{{{p_\text{d}}}}{{{\sigma ^2_\text{d}}}}{\left|{{{\mathop\mathbf{\hat{h}}\limits}^H_{t}}\mathop {\hat{\mathbf{w}}_t}\limits} \right|^2}\right) = {\log _2}\left(1 + \frac{{{ p_\text{d}}{{\left\| {{\mathbf{y}_{t}}} \right\|}^2} }}{{{\sigma^2 _\text{d}} {{ p}_{{\text{u}}}}}}\right).
\end{equation}

\subsubsection{Reflection Optimization}
We repeat performing channel estimation and active transmit beamforming for each RC in the codebook. After obtaining all objective function values, the optimal RC index can be obtained by selecting the best one maximizing the achievable rate from the designed codebook. The optimal index searching process can be formulated as
\begin{equation}
\label{6}
\hat{t} =\mathop {\arg}\limits_{{t}\in\mathcal{T}} \mathop {\max } {R_t} .
\end{equation}After obtaining the optimal index $\hat{t}$, the BS can determine the transmit beamforming vector $\mathbf{w}_{\hat{t}}$, and the RIS controller can configure the optimal RCs. Thus the codebook scheme obtains a suboptimal solution within a salable overhead.

\subsection{Environment-Aware Codebook Design} \label{mycodebook}
Existing codebook solutions including the random codebook \cite{ref14_WCL} and the sum Euclidean distance maximizing codebook \cite{an_equi} are non-adaptive, which generally results in worse performance in a limited training overhead. To solve this issue, we propose an adaptive codebook scheme by employing the statistical CSI relying on the LoS components of the channels. {In order to reduce the implementation complexity, we first select a reference antenna at the BS, and generate multiple virtual channels according to (\ref{generate_ch}), where the LoS and NLoS components of the virtual channels are generated by the steering vectors calculated by location information of the UE and a  Gaussian distribution. Assume that we have chosen the $m$-th BS antenna as the reference and generate  $T$ sets of virtual channels. Upon aligning all reflected channels with the direct channel for each virtual channel, the optimal phase shift of the $n$-th  element at the $t$-th training TS is given by 
\begin{align}
\label{7}
{\psi _{n,t}} =&\angle   {\left(\sqrt {{K_\text{r}}} {{h}^\text{LoS}_{\text{r},n}} + {\tilde{h}^{\text{NLoS}}_{\text{r},n,t}}\right)}
- \angle  {\left(\sqrt{{K_{\text{d}}}} {{{h}^\text{LoS}_{\text{d},m}}} + {\tilde{h}^{\text{NLoS}}_{\text{d},m,t}}\right)}   \notag\\ 
&- \angle \left(\sqrt {{K_{\text{g}}}} {{g}^\text{LoS}_{n,m}} + {\tilde{g}^{\text{NLoS}}_{n,m,t}}\right),  \quad {\forall}{n}\in\mathcal{N},\ {\forall}{t}\in\mathcal{T},
\end{align}where ${{h}^\text{LoS}_{\text{d},m}}$, ${{h}^\text{LoS}_{\text{r},n}}$, and ${{g}^\text{LoS}_{n,m}}$ 
are the LoS components of the $m$-th entry of $\mathbf{h}^\text{LoS}$, the $n$-th entry of ${\mathbf{h}^\text{LoS}_{\text{r}}}$, and the $(n,m)$-th entry of $\mathbf{G}^\text{LoS}$, respectively,  at the $t$-th TS. } ${\tilde{h}^{\text{NLoS}}_{\text{d},m,t}}$,  ${\tilde{h}^{\text{NLoS}}_{\text{r},n,t}}$, and ${\tilde{g}^{\text{NLoS}}_{n,m,t}}$ represent  the  $\text{NLoS}$ components of corresponding virtual channels which we generated off-line. The continuous RC of the $n$-th reflecting element at the $t$-th training TS can be denoted by ${ {\varphi}' _{n,t} }={ {e}^{ {j}{{\psi} _{n,t}}}}$. Next, we consider discrete phase shift ${ \hat{\varphi}_{n,t}}$  by minimizing  the quantization error, which can be denoted as 
\begin{align}
&\hat{\varphi}_{n,t} =\mathop {\arg}\limits_{ {{\varphi}_{s} } \in \mathscr{E} } {{  }}\mathop {\min } {\left| {{\varphi}'_{n,t}- {\varphi}_{s} } \right| }, \quad {\forall}{n}\in\mathcal{N}.
\end{align}Furthermore, the RC vector of the $t$-th training TS is denoted as ${{\boldsymbol{\phi} _t} = [{\hat{\varphi} _{1,t}}, {\hat{\varphi} _{2,t}},..., {\hat{\varphi} _{N,t}}]^T}$. Moreover, considering the fact that the generated ${\boldsymbol{\phi} _t}$ may conflict with previous ones, we generate a new one until $T$ diverse RC vectors are obtained.

\subsection{The Advantages of the Proposed Scheme} \label{Advantages}
{Next, we elaborate on several benefits of the proposed scheme. First, the proposed scheme's overhead is independent of $N$. Hence, the proposed scheme provides a beneficial trade-off between the system performance and training overhead.} In addition, the proposed codebook-based scheme  only needs feedback the optimal index of $\left\lceil {{{\log }_2}T} \right\rceil $ bits instead of existing counterparts requiring $Na$ bits to configure the RIS. Moreover, compared to the existing codebook designs \cite{ref14_WCL,an_equi}, the proposed scheme utilizes the statistical CSI and achieves substantial performance gain under the same overhead.

{Moreover, the complexity of channel estimation and reflection joint optimization of the AO algorithm are $2M\left(N+1\right)$ and $N_\text{iter}(2^{a}N+4MN+4M)$ in terms of the real-valued multiplications  \cite{ref14_WCL}, where $N_\text{iter}$ denotes the number of iterations for implementing the AO algorithm, $2^{a}N$, $4MN$, and $4M$ denote the complexity for optimizing reflection matrix $\mathbf{\Theta}$, calculating $\mathbf{h}^H$, and active beamforming $\mathbf{w}$ at each iteration.} By contrast, the complexity of channel estimation and the reflection optimization of the proposed scheme are $4MT$ and $(6M+8)T$, respectively, where $4M$, $6M+3$, and $5$  denote the complexity of estimating a composite channel, optimizing active beamforming, and calculating the achievable rate for each RC, respectively. The  complexity comparison will be demonstrated in Section \ref{Sim}.

\section{Theoretical Analysis} \label{the}
In this section, we analyze the scaling law of the average received power at the UE of the environment-aware codebook scheme. For simplicity, we assume $M=1$. Besides, the direct channel is blocked and the obtained CSI is accurate. We consider the Rician channel model for  $\mathbf{h}_\text{r}$ and $\mathbf{g}$. Specifically, $\mathbf{h}_r$ is generated following (\ref{generate_ch}), where $\left| h^{\text{LoS}}_{\text{r},n} \right|=1$ and $h^{\text{NLoS}}_{\text{r},n}  \sim \mathcal{CN}(0,1)$. The modeling of $\mathbf{g}$ is similar, and the Rician factor of the $\mathbf{g}$ is set to  $K_\text{g}\to \infty$. Moreover, the channels associated with different elements are i.i.d..  Rician fading with average power $\beta_r$ and $\beta_g$ for ${h}_{r,n}$ and ${g}_{n}$, respectively, where ${h}_{r,n}$ and ${g}_{n}$ denote the $n$-th entry of $\mathbf{h}_\text{r}$ and $\mathbf{g}$. Based on the above assumptions, we obtain the \textit {Proposition 1}. 

\emph{Proposition 1:} Assume ${h}_{\text{r},n}$ following the  Rician channel modeling with Rician factor of $K_\text{r}$, ${n}\in\mathcal{N}$. For $N>>1$, the signal power received at the UE is given by
\begin{equation}
\label{theory}
    p_\text{r} < {p_\text{d}}{\beta_\text{r}}{\beta_\text{g}}\left( \underbrace{{N^2} {{K^2_\text{1}}}}_{\left(\romannumeral1\right)}+\underbrace{N{{K^2_\text{2}}}\left(\log_{}{T} +C \right)}_{\left(\romannumeral2\right)} +\underbrace{N^2{K_\text{1}}{K_\text{2}}\sqrt{\pi}}_{\left(\romannumeral3\right)} \right),
\end{equation}where ${K_\text{1}}=\sqrt{{{{K_\text{r}}}}/ \left({{1 + {K_\text{r}}}} \right)}$, ${K_\text{2}}=\sqrt{{1}/\left({{1 + {K_\text{r}}}}\right)}$, and $C\approx0.5772$ is the Euler-Mascheroni constant.

\emph{Proof:} By applying the above conditions, the signal power received at the UE is denoted by $p_\text{r}={p_\text{d}}\mathbb{E}\left\{ \mathop {\max }\limits_{{t}\in\mathcal{T}}  \left|{\boldsymbol{\phi}^T _t}\mathbf{d} \right|^2 \right\}$, which can be further expressed as
\begin{align}
\label{procc}
 &\quad\ {p_\text{d}}\mathbb{E}\left\{ \mathop {\max }\limits_{{t}\in\mathcal{T}}  \left|{\boldsymbol{\phi}^T _t}\mathbf{d} \right|^2 \right\} \notag\\
        &= {p_\text{d} {\beta_r}{\beta_g}}\mathbb{E}\left\{  \mathop {\max }\limits_{{t}\in\mathcal{T}}  { \left| {{\sum\limits_{n = 1}^N{\hat{\phi}_{n,t}}( {K_\text{1}} {{{h}^{\text{LoS}}_{\text{r},n}} +  {K_\text{2}} {{h}^{\text{NLoS}}_{\text{r},n}})^*}}}{g_n} \right|^2   } \right\}  \notag\\
        &={p_\text{d}{\beta_r}{\beta_g}}\mathbb{E} \left\{ \mathop {\max }\limits_{{t}\in\mathcal{T}} \left\{     {\left|{  \sum\limits_{n = 1}^N{\hat{\phi}_{n,t}} {K_\text{1}} ({{h}^{\text{LoS}}_{\text{r},n}})^* {g_n}}  \right|}^2  \right.\right.\notag\\
     & + \left.\left. { \left| { \sum\limits_{n = 1}^N{\hat{\phi}_{n,t}} {K_\text{2}}
        ({{h}^{\text{NLoS}}_{\text{r},n}})^*{g_n} } \right|^2}+2\text{Re} \left\{  {\sum\limits_{n = 1}^N \left({\hat{\phi}_{n,t}} {K_\text{1}} ({{h}^{\text{LoS}}_{\text{r},n}})^*{g_n}\right)^*}   \right.\right.\right.\notag\\ 
        &\left.\left.\left.   \times{\sum\limits_{n = 1}^N\left({\hat{\phi}_{n,t}} {K_\text{2}} ({{h}^{\text{NLoS}}_{\text{r},n}})^*{g_n}\right)} \right\}  \right\} \right\}, 
\end{align}where $\mathbf{d}$ denotes the  cascaded BS-RIS-UE channel of  $\mathbf{g}$ and $\mathbf{h}_\text{r}$, while ${\hat{\phi}_{n,t}}$ is obtained in Section \ref{mycodebook}. Since obtaining a closed solution for (\ref{procc}) is non-trivial, we next derive a theoretical upper bound by scaling  three entries in (\ref{procc}). First, by invoking the Lindeberg-Lévy central limit theorem \cite{ref14_WCL}, we have $ \sum\limits_{n = 1}^N  {\hat{\phi}_{n,t}}{K_\text{1}}( {{ {h}^{\text{LoS}}_{\text{r},n} })^*}{g_n} \sim \mathcal{CN}(0,N{K^2_\text{1}})$ as $N\to \infty$. Furthermore, we assume that the dominant LoS components of  reflected channels are aligned in the first entry of (\ref{procc}),  then we obtain  $\left(\romannumeral1\right)$ of (\ref{theory})  by applying (31) in \cite{ref1_AO}. Second, we assume that the NLoS components of  reflected channels to be dominant and adopt multiple sets of random RCs to optimize the NLoS components of channels in the second entry of  (\ref{procc}), thus we  obtain $\left(\romannumeral2\right)$ of (\ref{theory}) according to (13) in \cite{ref14_WCL}. Finally, we consider both the LoS and NLoS components of reflected channels in the third entry of (\ref{procc}). Due to the fact that $\left| {h}^{\text{NLoS}}_{\text{r},n} \right|$ follows Rayleigh  distribution with mean value of $\sqrt{\pi}/2$, we have an upper bound of $2\sum\limits_{n = 1}^N\left( {K_\text{1}}\left|{h}^{\text{LoS}}_{\text{r},n} \right| \left| {g_n} \right|\right) \sum\limits_{n = 1}^N \left({K_\text{2}}\left|{h}^{\text{NLoS}}_{\text{r},n} \right| \left| {g_n} \right| \right)$, based on which we can readily obtain  $\left(\romannumeral3\right)$ of (\ref{theory}). $\hfill\blacksquare$

It is noted that the performance of the proposed scheme is highly dependent on the channel structure. Specifically, when $K_\text{1} \to 1$, i.e., the reflected channel is equivalent to the virtual LoS channel, we have $p_\text{r}\to{p_\text{d}}{\beta_\text{r}}{\beta_\text{g}}{N^2} {K^2_\text{1}}$, which characterizes the quadratic power scaling law versus the number of RIS elements \cite{ref1_AO}. When  considering the Rayleigh channel for $\mathbf{h}_r$, i.e., $K_\text{2}\to 1$, we have $p_\text{r}\to{p_\text{d}}{\beta_\text{r}}{\beta_\text{g}}{N{{K^2_\text{2}}}\left(\log_{}{T} +C \right)}$, which is consistent with the conclusion of \cite{an_oujld}. Furthermore, when  $T=1$, we have  $p_\text{r}={p_\text{d}}{\beta_\text{r}}{\beta_\text{g}}{N{{K^2_\text{2}}}C}$. Moreover, when considering the maximum overhead of $T_\text{max}=2^{aN}$, we have $p_\text{r}\to{a}{p_\text{d}}{\beta_\text{r}}{\beta_\text{g}}{N^2{{K^2_\text{2}}}}$.

\section{Simulation Results} \label{Sim}
In this section, we provide simulation results to verify the performance of our proposed scheme. We consider a 3D Cartesian coordinate system, where the antenna array at the BS is modeled by a ULA and  deployed on the y-axis with antenna spacing of ${d_\text{BS}} = \lambda /2$. The UE is located on the $x$-axis. In addition, we assume that the RIS  is deployed as a UPA  on the $x-z$ plane with $10 \times 10$ array structure and element spacing of ${d_\text{R}} = \lambda /8$. { Moreover, if not specified, we consider 1 bit phase quantization, i.e., $\mathscr{E} =\{{e^{j0}}, {e^{j\pi }}\}$, which is common in the practical design \cite{ref14_WCL}.} The coordinates of the BS, RIS, and UE are $(0,0,{h_\text{BS}})$, $({d_\text{BU}}, {d_\text{RU}}, {h_\text{R}})$, and $({d_\text{BU}},0,0)$, respectively, where the distance from the BS and RIS to UE are ${d_\text{BU}} = 100$ m and ${d_\text{RU}} = 6$ m on $x-y$ plane, respectively, the height of both the BS and RIS  are ${h_\text{BS}}={h_\text{R}}=5$ m.  Thus,  The path loss of each channel is modeled as $\beta = {C_\text{0}}{(d/{d_\text{0}})^{ - \alpha }}$, where ${C_\text{0}}=-20$ dB denotes the path loss at the reference distance of ${d_\text{0}}=1$ m, $d$ denotes the distance of the link. Moreover, the path loss factor of $\mathbf{G}$, ${\mathbf{h}_\text{r}}$, and ${{\mathbf{h}_\text{d}}}$ are set to ${\alpha _\text{g}} = 2.4$, ${\alpha _\text{r}} = 2.5$, and ${\alpha _\text{d}} = 3.5$, respectively. In addition, the Rician factors of the $\mathbf{G}$, ${\mathbf{h}_\text{r}}$, and  ${{\mathbf{h}_\text{d}}}$ are  set to ${K_\text{g}} = 4$ dB, ${K_\text{r}} = 3$ dB, and ${K_\text{d}} =-3$ dB, respectively.  Moreover, we consider  the  transmit power at the BS is ${p_\text{d}} = 40$ dBm,  the average noise power at the BS and the UE are ${\sigma^2_\text{u}} = - 110$ dBm and ${\sigma^2 _\text{d}} = -90$ dBm, respectively. Moreover, all results are obtained by averaging over 1,000 independent experiments.

\begin{figure}[!t]
\centering
\subfloat[Under perfect CSI.]{\includegraphics[width=3.5cm]{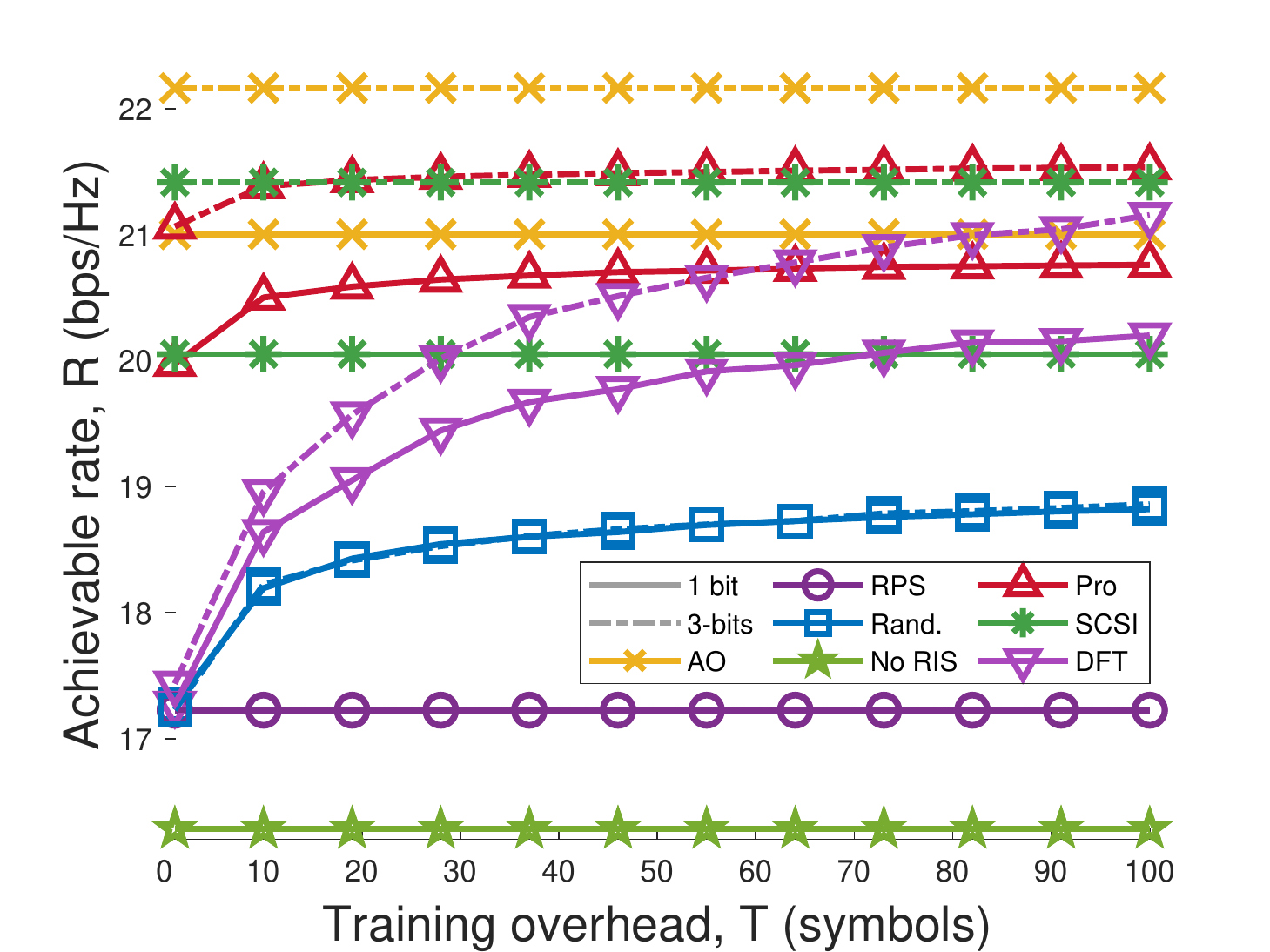}}\qquad
\subfloat[Under imperfect CSI. ($p_\text{u}=0$ dB).]{\includegraphics[width=3.5cm]{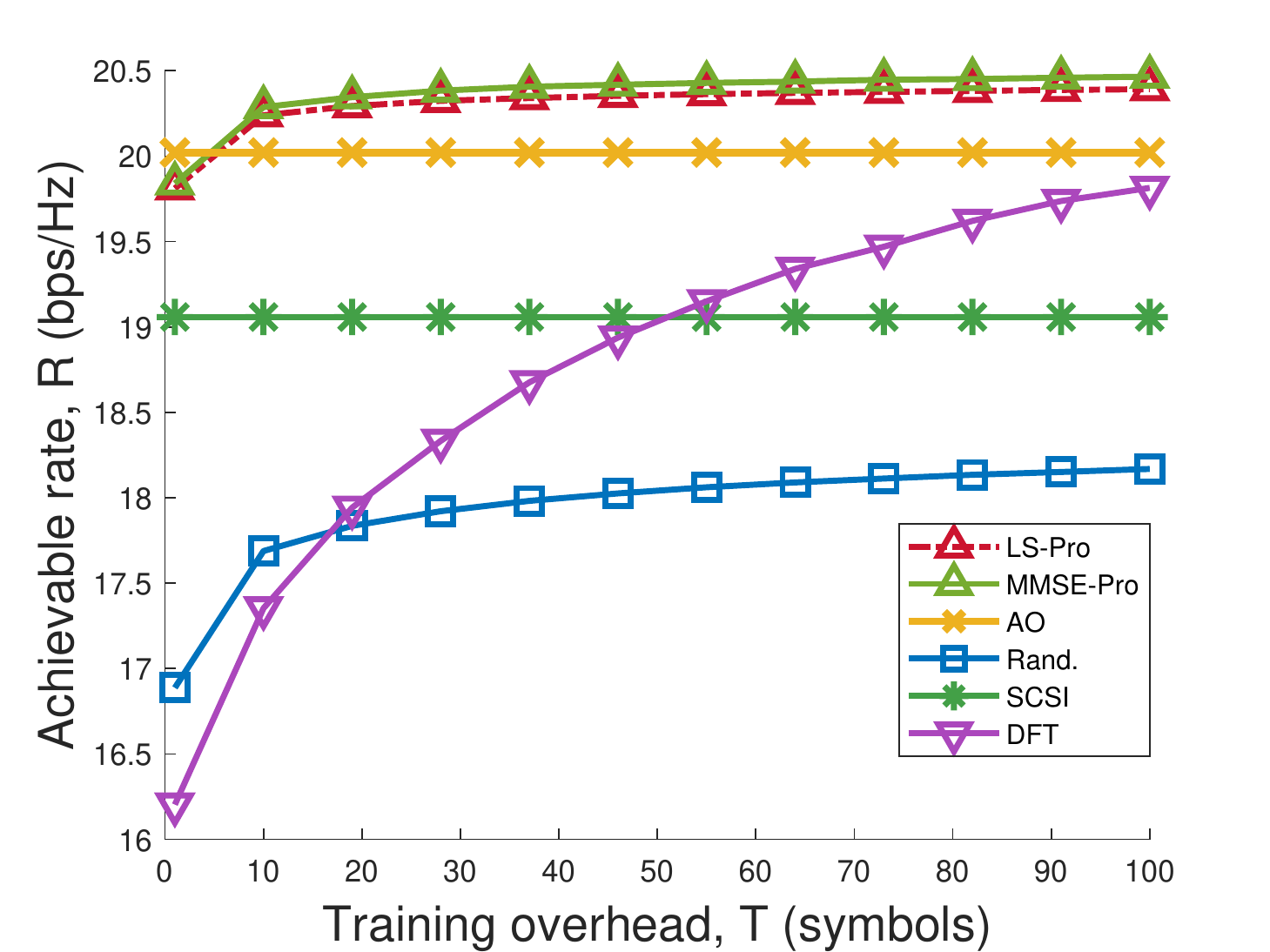}}\qquad
\caption{The achievable rates versus training overhead ($N=100$, $M=8$, $N_\text{iter}=3$). }
\vspace{-0.9cm} 
\label{fig_3}
\end{figure}

{ As shown in Fig. \ref{fig_3}(a), we evaluate the achievable rate versus different training overhead, where we consider six benchmark schemes, including the AO algorithm \cite{ref1_AO}, random phase shift (RPS) scheme \cite{ref14_WCL}, random codebook  (\emph{Rand.}) scheme \cite{ref14_WCL}, plain MISO without RIS  \cite{ref1_AO}, and statistical CSI (SCSI)-based scheme \cite{Gua_HY} without selecting a reference antenna with $T=1$. Note that the achievable rate of the proposed scheme, AO, and the SCSI scheme can be improved as the number of  quantization bits b increases, while the random codebook scheme hardly attain any performance gain \cite{ref14_WCL}. Moreover, the proposed scheme has a moderate rate loss compared to the AO algorithm under the perfect CSI, which, however, can be gradually improved with the increase of $T$. In addition, the system's achievable rate of the proposed scheme benefiting from the statistical CSI outperforms the random codebook and DFT scheme at $T=1$. 
Fig. \ref{fig_3}(b) compares the achievable rate of different schemes and channel estimation techniques under imperfect CSI. It is worth noting that the rate of the proposed scheme even outperforms the  AO scheme in the presence of channel estimation errors. However, the AO algorithm still requires 101 training TSs to attain such inaccurate CSI. As a result, our scheme is more competitive in the face of imperfect CSI.}

\begin{figure}[!t]
\centering
\subfloat[Under perfect CSI.] {\includegraphics[width=3.5cm]{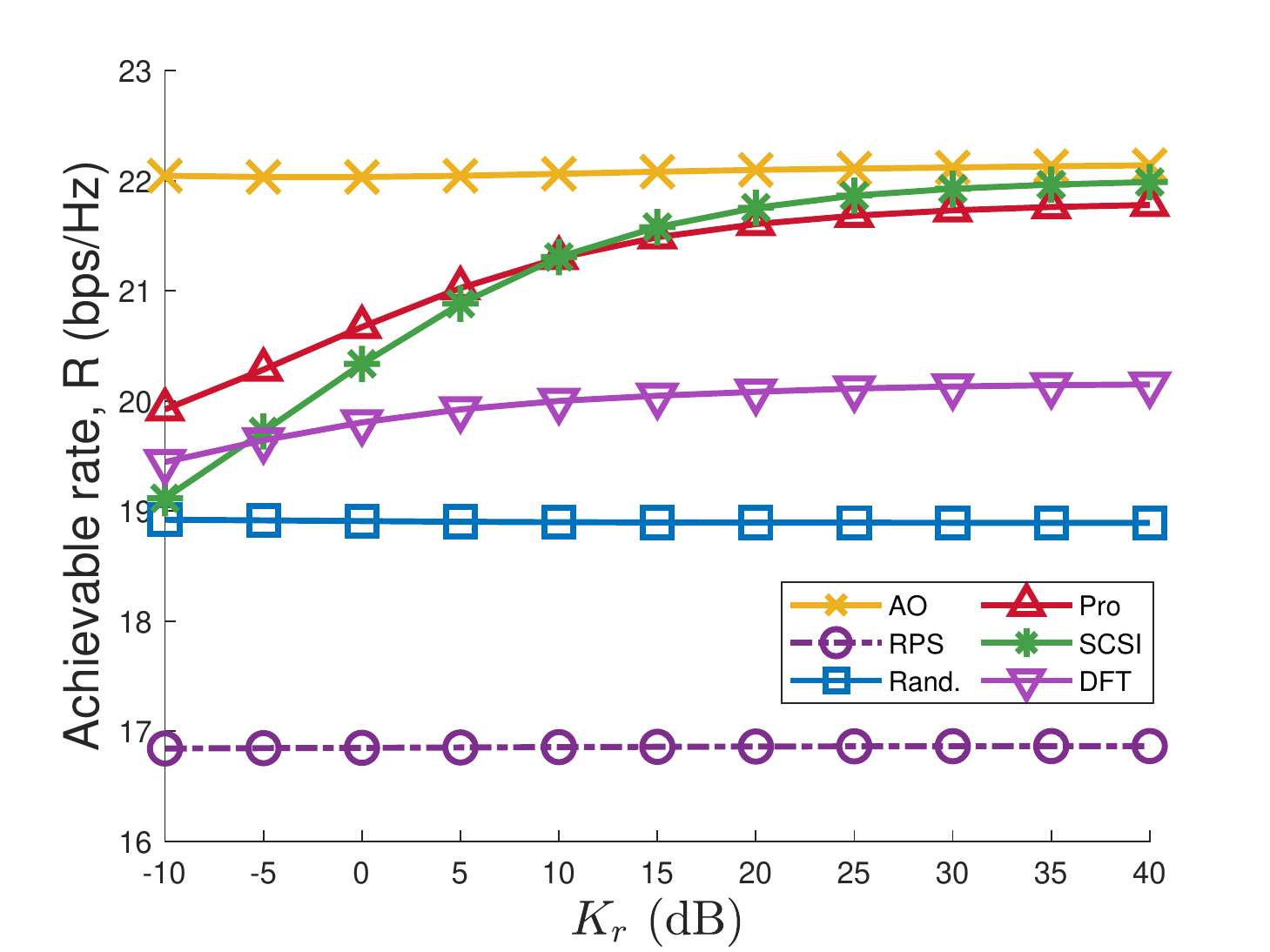}}\qquad
\subfloat[Under imperfect CSI ($p_\text{u}=-20$ dB).]{\includegraphics[width=3.5cm]{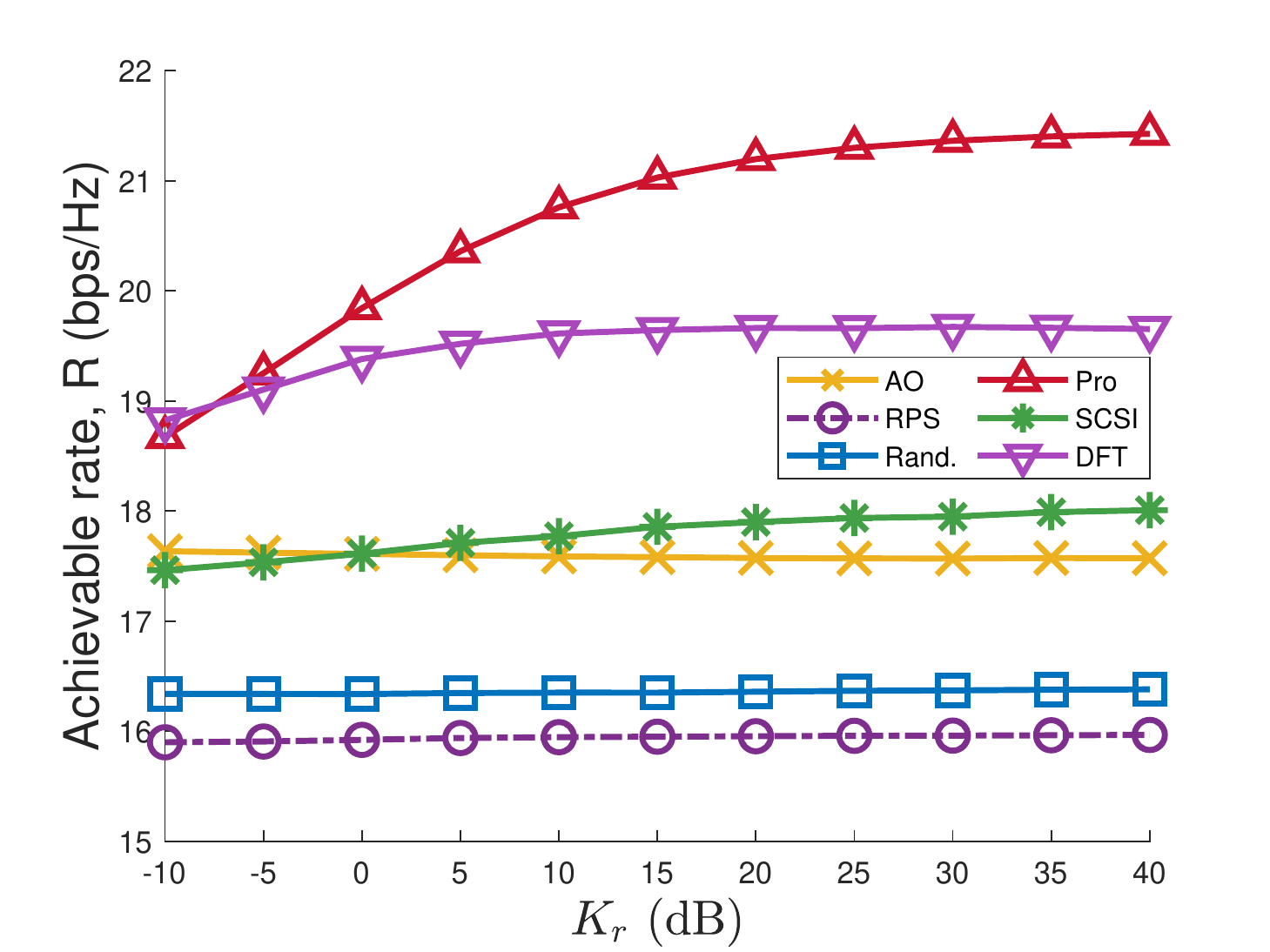}}\qquad
\caption{The achievable rates versus Rician factor ${K_\text{r}}$ ($N=100$, $M=8$, $N_\text{iter}=3$, $T=50$).}
\vspace{-0.85cm} 
\label{fig_4}
\end{figure}

{As shown in Fig. \ref{fig_4}(a), we consider ${K_\text{g}} = {K_\text{d}} \to \infty$ and analyse the influence of ${K_\text{r}}$ on the achievable rate. Observe from Fig. \ref{fig_4}(a) that the achievable rate of the proposed, DFT, and SCSI schemes gradually increase with the value of $K_r$. The proposed  and SCSI schemes  almost achieve the same achievable rate with the AO algorithm under perfect CSI at ${K_\text{r}}=30$ dB. Moreover, the proposed scheme relying on an environment-aware codebook significantly outperforms the random codebook, RPS, and DFT schemes.} Fig. \ref{fig_4}(b) considers the imperfect CSI by taking ${p_\text{u}} =-20$ dBm. Again, the proposed scheme outperforms the AO algorithm in the face of severe channel estimation errors. As shown in Fig. \ref{fig_4}, the achievable rate of the proposed scheme improves as ${K_\text{r}}$ increases, while the other benchmark schemes are not sensitive to  ${K_{\text{r}}}$. In a nutshell, our scheme utilizes the statistical CSI to attain better performance.   

\begin{figure}[!t]
\centering
\subfloat[]{\includegraphics[width=3.5cm]{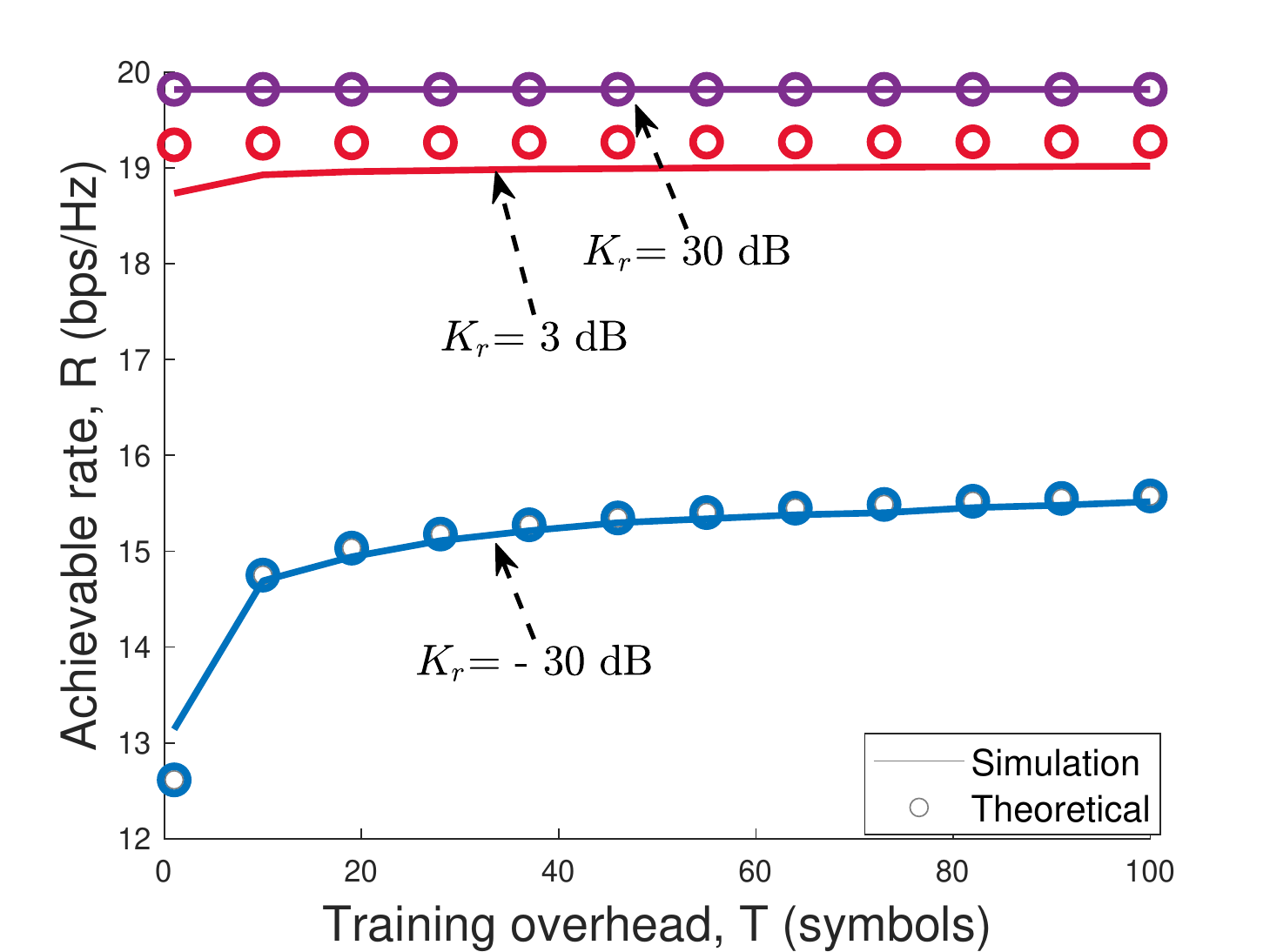}}\qquad
\subfloat[]{\includegraphics[width=3.5cm]{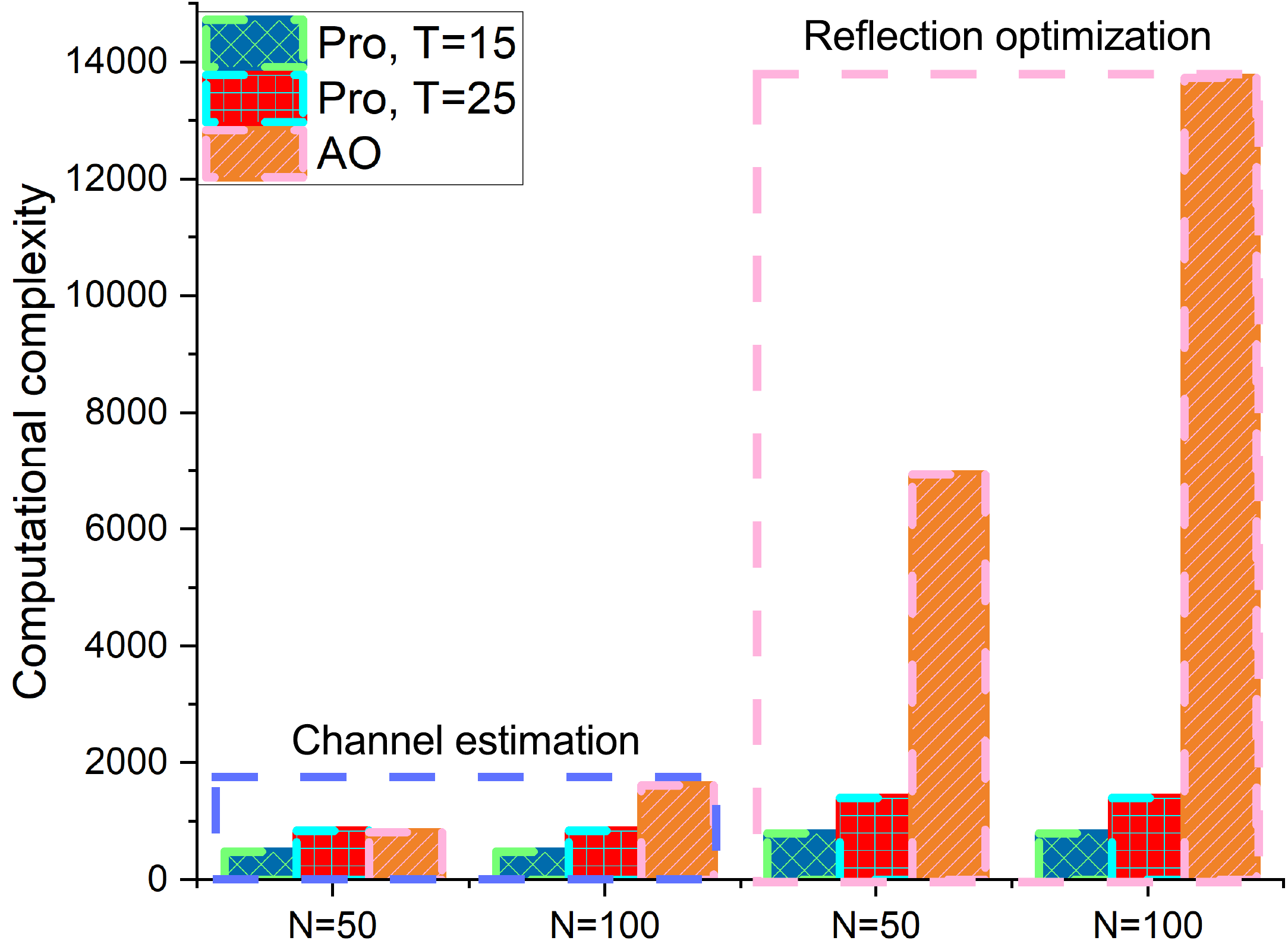}}\qquad
\caption{(a) The achievable rate comparison of theoretical and simulation results under different ${K_\text{r}}$ ($N=100$, $M=1$). (b) The complexity comparison of the proposed scheme and the AO algorithm ($M=8$, $N_\text{iter}=4$).}
\vspace{-0.8cm} %
\label{fig_5}
\end{figure}

 Fig. \ref{fig_5}(a) verifies our analytical results in Section \ref{the},  where we set $K_r=-30$ dB, $3$ dB, and $30$ dB, respectively. Note that the theoretical results serve a tight upper bound of the simulation results under all setups. Finally, Fig. \ref{fig_5}(b) compares the complexity of the AO algorithm and the proposed scheme. In sharp contrast to the AO algorithm, the complexity of the proposed scheme is independent of the number of reflecting elements $N$. In addition, with the increase of $T$, the complexity of the proposed scheme will increase, which, however, is still less than the AO algorithm. 

\section{Conclusions}
In this letter, we proposed an environment-aware codebook scheme by exploiting the statistical CSI for the RIS-assisted MISO system. We perform reflection optimization for each RC in the pre-designed codebook and obtain the best RC by maximizing the objective function. Furthermore, we analyzed the theoretical achievable rate of the proposed scheme. Moreover, simulation results demonstrated that our proposed scheme attained performance benefit under imperfect CSI, albeit its low implementation complexity and  training overhead. {Note that the proposed scheme can be readily extended to multiple-input multiple-output and multi-user MISO scenarios by exploiting statistical CSI-based joint optimization methods.}  
\bibliographystyle{IEEEtran}
\bibliography{IEEEabrv,mylib}

\begin{thebibliography}{10}
\providecommand{\url}[1]{#1}
\csname url@samestyle\endcsname
\providecommand{\newblock}{\relax}
\providecommand{\bibinfo}[2]{#2}
\providecommand{\BIBentrySTDinterwordspacing}{\spaceskip=0pt\relax}
\providecommand{\BIBentryALTinterwordstretchfactor}{4}
\providecommand{\BIBentryALTinterwordspacing}{\spaceskip=\fontdimen2\font plus
\BIBentryALTinterwordstretchfactor\fontdimen3\font minus
  \fontdimen4\font\relax}
\providecommand{\BIBforeignlanguage}[2]{{%
\expandafter\ifx\csname l@#1\endcsname\relax
\typeout{** WARNING: IEEEtran.bst: No hyphenation pattern has been}%
\typeout{** loaded for the language `#1'. Using the pattern for}%
\typeout{** the default language instead.}%
\else
\language=\csname l@#1\endcsname
\fi
#2}}
\providecommand{\BIBdecl}{\relax}
\BIBdecl

\bibitem{ref1_AO}
Q.~Wu and R.~Zhang, ``Intelligent reflecting surface enhanced wireless network
  via joint active and passive beamforming,'' \emph{IEEE Trans. Wireless
  Commun.}, vol.~18, no.~11, pp. 5394--5409, Nov. 2019.

\bibitem{an_equi}
J.~An, C.~Xu, L.~Wang, Y.~Liu, L.~Gan, and L.~Hanzo, ``Joint training of the
  superimposed direct and reflected links in reconfigurable intelligent surface
  assisted multiuser communications,'' \emph{IEEE Trans. Green Commun. Netw.},
  vol.~6, no.~2, pp. 739--754, Jun. 2022.

\bibitem{ref2_C1}
C.~You, B.~Zheng, and R.~Zhang, ``Channel estimation and passive beamforming
  for intelligent reflecting surface: Discrete phase shift and progressive
  refinement,'' \emph{IEEE J. Sel. Areas Commun.}, vol.~38, no.~11, pp.
  2604--2620, Nov. 2020.

\bibitem{xr3}
C.~Huang, A.~Zappone, G.~C. Alexandropoulos, M.~Debbah, and C.~Yuen,
  ``Reconfigurable intelligent surfaces for energy efficiency in wireless
  communication,'' \emph{IEEE Trans. Wireless Commun.}, vol.~18, no.~8, pp.
  4157--4170, Aug. 2019.

\bibitem{Gua_HY}
H.~Guo and V.~K. Lau, ``Uplink cascaded channel estimation for intelligent
  reflecting surface assisted multiuser {MISO} systems,'' \emph{IEEE Trans.
  Signal Process.}, vol.~70, pp. 3964--3977, Jul. 2022.

\bibitem{zhangSB}
S.~Zhang, S.~Zhang, F.~Gao, J.~Ma, and O.~A. Dobre, ``Deep learning optimized
  sparse antenna activation for reconfigurable intelligent surface assisted
  communication,'' \emph{IEEE Trans. Commun.}, vol.~69, no.~10, pp. 6691--6705,
  Jul. 2021.

\bibitem{xc3}
W.~Xie, J.~Xiao, P.~Zhu, C.~Yu, and L.~Yang, ``Deep compressed sensing-based
  cascaded channel estimation for {RIS}-aided communication systems,''
  \emph{IEEE Wireless Commun. Lett.}, vol.~11, no.~4, pp. 846--850, Apr. 2022.

\bibitem{ref8_R3}
H.~Li, W.~Cai, Y.~Liu, M.~Li, Q.~Liu, and Q.~Wu, ``Intelligent reflecting
  surface enhanced wideband {MIMO}-{OFDM} communications: From practical model
  to reflection optimization,'' \emph{IEEE Trans. Commun.}, vol.~69, no.~7, pp.
  4807--4820, Jul. 2021.

\bibitem{STA_CSI}
Y.~Han, W.~Tang, S.~Jin, C.-K. Wen, and X.~Ma, ``Large intelligent
  surface-assisted wireless communication exploiting statistical {CSI},''
  \emph{IEEE Trans. Veh. Technol.}, vol.~68, no.~8, pp. 8238--8242, Aug. 2019.

\bibitem{xr1}
W.~Yan, G.~Sun, W.~Hao, Z.~Zhu, Z.~Chu, and P.~Xiao, ``Machine learning-based
  beamforming design for millimeter wave {IRS} communications with discrete
  phase shifters,'' \emph{IEEE Wireless Commun. Lett.}, vol.~11, no.~12, pp.
  2467--2471, Mar. 2022.

\bibitem{ref14_WCL}
J.~An and L.~Gan, ``The low-complexity design and optimal training overhead for
  {IRS}-assisted {MISO} systems,'' \emph{IEEE Wireless Commun. Lett.}, vol.~10,
  no.~8, pp. 1820--1824, Aug. 2021.

\bibitem{an_oujld}
J.~An \emph{et~al.}, ``Low-complexity channel estimation and passive
  beamforming for {RIS}-assisted {MIMO} systems relying on discrete phase
  shifts,'' \emph{IEEE Trans. Commun.}, vol.~70, no.~2, pp. 1245--1260, Feb.
  2022.

\bibitem{ref10_M2}
W.~Chen, C.-K. Wen, X.~Li, and S.~Jin, ``Adaptive bit partitioning for
  reconfigurable intelligent surface assisted {FDD} systems with limited
  feedback,'' \emph{IEEE Trans. Wireless Commun.}, vol.~21, no.~4, pp.
  2488--2505, Apr. 2022.

\bibitem{kim}
J.~Kim, S.~Hosseinalipour, A.~C. Marcum, T.~Kim, D.~J. Love, and C.~G. Brinton,
  ``Learning-based adaptive {IRS} control with limited feedback codebooks,''
  \emph{IEEE Trans. Wireless Commun.}, vol.~21, no.~11, pp. 9566--9581, Jun.
  2022.

\bibitem{tangWKl}
W.~Tang, X.~Chen, M.~Z. Chen, J.~Y. Dai, Y.~Han, S.~Jin, Q.~Cheng, G.~Y. Li,
  and T.~J. Cui, ``On channel reciprocity in reconfigurable intelligent surface
  assisted wireless networks,'' \emph{IEEE Wireless Commun.}, vol.~28, no.~6,
  pp. 94--101, Dec. 2021.

\end{thebibliography}
\end{document}